\documentclass[reqno,a4paper,final]{amsart}
\usepackage[left=3cm,right=3cm,top=2cm,bottom=2cm,includeheadfoot]{geometry}
\usepackage{amsmath}
\usepackage{amsfonts}
\usepackage{amssymb}
\usepackage{amsfonts}
\usepackage{MnSymbol}
\usepackage{amsmath}
\usepackage{cancel}
\usepackage{graphicx}
\usepackage[numbers]{natbib}
\usepackage{longtable}
\usepackage{xcolor}
\usepackage{appendix}

\usepackage[ansinew]{inputenc}
\usepackage{amsthm}
\usepackage[arrow, matrix, curve]{xy}
\usepackage{hyperref}
\hypersetup{%
  pdftitle   = {(M-theory-)Killing spinors on symmetric spaces},
  pdfkeywords = {supergravity, Killing spinors, isometries,
    homogeneity, supersymmetry, symmetric spaces, M-theory},
  pdfauthor  = {Noel Hustler, Andree Lischewski},
  pdfcreator = {\LaTeX\ with package \flqq hyperref\frqq}
}
\newtheoremstyle{style}   
  {0.5cm}                 
  {0.5cm}                 
  {}                         
  {}                         
  {\normalfont\bfseries}  
  {\normalfont }{   }
  {}
\theoremstyle{style}

\newtheorem{definition}{Definition}\numberwithin{definition}{section}
\newtheorem{remark}[definition]{Remark}
\newtheorem{Lemma}[definition]{Lemma}
\newtheorem{example}[definition]{Example}
\newcommand{\ph}{\varphi}
\newcommand{\R}{\mathbb{R}}

\newcommand{\C}{\mathbb{C}}

\newcommand{\Pe}{\mathcal{P}}

\newcommand{\g}{\mathfrak{g}}

\allowdisplaybreaks[1]
\begin{document}
\title{(M-theory-)Killing spinors on symmetric spaces}
\author[Noel Hustler]{Noel Hustler}
\address[Noel Hustler]{Maxwell and Tait Institutes, School of Mathematics, University of Edinburgh, King's Buildings, Edinburgh EH9 3JZ, Scotland, United Kingdom}
\email{n.hustler@ed.ac.uk}

\author[Andree Lischewski]{Andree Lischewski}
\address[Andree Lischewski]{Humboldt-Universit\"at zu Berlin, Institut f\"ur Mathematik\\
Rudower Chaussee 25, Room 1.310, D12489 Berlin, Germany}
\email{lischews@mathematik.hu-berlin.de}

\thanks{EMPG-15-03}
\begin{abstract} 
We show how the theory of invariant principal bundle connections for reductive homogeneous spaces can be applied to determine the holonomy of generalised Killing spinor covariant derivatives of the form $D= \nabla + \Omega$  in a purely algebraic and algorithmic way, where $\Omega : TM \rightarrow \Lambda^*(TM)$ is a left-invariant homomorphism. Specialising this to the case of symmetric $M-$theory backgrounds (i.e.\ $(M,g,F)$ with $(M,g)$ a symmetric space and $F$ an invariant closed 4-form), we derive several criteria for such a background to preserve some supersymmetry and consequently find all supersymmetric symmetric $M-$theory backgrounds.
\end{abstract}
\maketitle

\tableofcontents

\section{Introduction}

The strong homogeneity theorem for supergravity backgrounds\cite{hom1,hom2,hom3} tells us that supergravity backgrounds preserving more than half of the maximum supersymmetry are homogeneous and moreover that this homogeneity is a direct consequence of the supersymmetries.  The homogeneity theorem means that a classification of homogeneous backgrounds would automatically give us a classification of backgrounds preserving most of the maximum supersymmetry.

However, one would think that a programme for classifying homogeneous supergravity backgrounds would necessarily require first classifying homogeneous Lorentzian manifolds in the relevant dimension -- and this would seem very difficult.
We are compelled to first attack the lowest-hanging fruit and the clear candidate in this case are the symmetric backgrounds.

A classification of all symmetric M-theory backgrounds was made in \cite{jose} (and additionally for symmetric Type IIB backgrounds in \cite{jose3}) using \'Elie Cartan's classification of irreducible Riemannian symmetric spaces (e.g. \cite{Helgason2001}) and Cahen and Wallach's classification of indecomposable Lorentzian symmetric spaces \cite{Cahen1,Cahen2}.  However this classification dealt with backgrounds at the bosonic level and said nothing about which backgrounds are supersymmetric.  In this paper we build on this result by determining which symmetric M-theory backgrounds are supersymmetric and if so what fraction of the maximum supersymmetry is preserved.  This represents the next step in classifying homogeneous M-theory backgrounds and concludes the classification of supersymmetric symmetric M-theory backgrounds.

In Section \ref{prelim} we will first present some basic material about (reductive) homogeneous spaces.  In Section \ref{pura} we describe how to algorithmically construct the holonomy of (left-invariant) spinor connections.  Section \ref{symback} demonstrates how to apply this technique in order to determine the supersymmetry of symmetric M-theory backgrounds via the holonomy representation of the superconnection.  Finally, in Section \ref{exclude} we run through the backgrounds from \cite{jose} and determine which are supersymmetric.

\section{Preliminaries}
\label{prelim}
We present basic facts about (reductive) homogeneous spaces, following \cite{aa}. In this article, let $M$ be a homogeneous space, i.e. $M=H/K$ for some Lie group $H$ and closed subgroup $K$. Let $\pi^H:H \rightarrow H/K$ denote the canonical projection. Further, let $L_h:H \rightarrow H$ and $l_h: H/K \rightarrow H/K$ be the left translations by $h \in H$. They are related by
\begin{align*}
\pi^H \circ L_h = l_h \circ \pi^H.
\end{align*}
Evaluating the differential $d\pi^H_e$ at the neutral element $e \in H$ yields that ker $d \pi_e^H = \mathfrak{k}$, and since $d\pi^H$ is onto, this gives rise to an isomorphism
\begin{align*}
\mathfrak{h} / \mathfrak{k} \cong T_{eK} \left(H/K \right).
\end{align*}
From now on we shall in addition assume that $H/K$ is \textit{reductive}, i.e. there exists a subspace $\mathfrak{n}$ of $\mathfrak{h}$ such that
\begin{align}
\mathfrak{h} = \mathfrak{k} \oplus \mathfrak{n} \text{ and } \left[\mathfrak{k},\mathfrak{n} \right] \subset \mathfrak{n}. \label{redu}
\end{align}
The \textit{isotropy representation} of $H / K$ is the homomorphism
\begin{align}
Ad^{H/K} : K \rightarrow GL\left( \mathfrak{n} \cong T_{eK}\left( H/K \right) \right),\text{ }k \mapsto (dl_k)_{eK}, \label{imp}
\end{align}
and it is equivalent to the adjoint representation of $K$ in $ \mathfrak{n}$, i.e. the diagram
\begin{align*}
\begin{xy}
  \xymatrix{
      \mathfrak{n}  \ar@{->}[r]^{Ad^H(k)} \ar[d]^{d \pi^H_e|\mathfrak{n}}    &   \mathfrak{n} \ar[d]^{d \pi^H_e|\mathfrak{n}}  \\
      T_{eK}\left(H/K \right) \ar@{->}[r]^{(dl_k)_{eK}}   &   T_{eK}\left(H/K \right)   \\
  }
\end{xy}
\end{align*}

commutes, where the upper horizontal map is obtained from restriction of $Ad^H(k):\mathfrak{h} \rightarrow \mathfrak{h}$.\\

Let us now turn to pseudo-Riemannian homogeneous spaces. A metric $g$ on $H/K$ is called $H-$\textit{invariant} if for each $h \in H$ the diffeomorphism  $l_h$ is an isometry wrt. $g$. It is well known that there is a one-to-one correspondence between $H-$invariant Riemannian metrics on $H/K$ and $Ad^{H/K}$-invariant scalar products on $\mathfrak{n}$. More generally, $H-$invariant tensor fields of type $(p,q)$ on $H/K$ correspond to $Ad^{H/K}$ invariant tensors of the same type on $\mathfrak{n}$, where the correspondence is given by evaluating the tensor field at the origin $eK \in H/K$. Note that in case of a $H-$invariant metric, the map (\ref{imp}) takes values in $O(\mathfrak{n}, \langle \cdot, \cdot \rangle_{\mathfrak{n}}),$ where from now on we denote by $\langle \cdot, \cdot \rangle_{\mathfrak{n}}$ the invariant inner product on $\mathfrak{n}$ corresponding to $g$.

In this paper, we shall mostly be concerned with \textit{(locally) symmetric spaces}. $(M,g)$ is locally symmetric if for every $x \in M$ there exists a normal neighbourhood $U$ of $x$ such that 
\begin{align*}
exp_x \circ (-Id_x) \circ exp_x^{-1}:U \rightarrow M
\end{align*}
is a local isometry. Equivalently, $(M,g)$ is locally symmetric iff $\nabla^g R^g = 0$. 
A symmetric space can be given the structure of a reductive homogeneous space which additionally satisfies 
\begin{align*}
\left[ \mathfrak{n}, \mathfrak{n} \right] \subset \mathfrak{k}.
\end{align*}
We now turn to spin structures on pseudo-Riemannian reductive homogeneous spaces. Let us assume that $(M=H/K,g)$ is space-and-time oriented\footnote{This assumption is only needed to make use of the natural inner product on the spinor bundle when considering the Killing superalgebra later. In all other cases it suffices that $M$ is oriented with the obvious modifications.}. In this case the bundle $\mathcal{P}^g \rightarrow M$ of oriented frames of $(M,g)$ can be reduced to structure group $SO^+(p,q)$\footnote{In this text, the $+$ denotes the identity component of a Lie group.}. Moreover, one has that 
\begin{align*}
\mathcal{P}^g \cong H \times_{Ad^{H/K}} SO^+(\mathfrak{n}),
\end{align*}
and
\begin{align}
\mathcal{P}^g \times_{SO^+(p,q)}\R^n \cong TM \cong H\times_{Ad^{H/K}} \mathfrak{n}, \label{tb}
\end{align}
where the latter isomorphism is given by $dl_h d\pi_e X \mapsto [h,X]$ for $X \in T_{eK} \left(H/K \right)$.\\
Any lift of the isotropy representation $Ad^{H/K}$ to $Spin^+(\mathfrak{n})$, i.e. any map $\widetilde{Ad}^{H/K}:K \rightarrow Spin^+(\mathfrak{n})$ such that the diagram
\begin{align*}
\begin{xy}
  \xymatrix{
      Spin^+(\mathfrak{n})    \ar[rd]^{\lambda}    &     \\
      K \ar@{->}[r]^{Ad^{H/K}} \ar[u]^{\widetilde{Ad}^H(k)} &   SO^+(\mathfrak{n})   \\
  }
\end{xy}
\end{align*}

commutes, allows us to fix a \textit{homogeneous spin structure} $(\mathcal{Q}^g=H \times_{\widetilde{Ad}^{H/K}}Spin^+(\mathfrak{n}),f^g)$ of $(M,g)$ (cf. \cite{bfkg}), where $f^g:\mathcal{Q}^g \rightarrow \mathcal{P}^g$ is simply the double covering $\lambda:Spin(p,q) \rightarrow SO(p,q)$ in the second factor. From now on we shall always assume that $(M,g)$ admits a homogeneous spin structure and think of this structure as being fixed. \\
\newline
Finally, we follow \cite{kn1,kn2} to observe that the natural $H-$left action on a reductive homogeneous space induces a $H-$left action $L$ on \textit{homogeneous fibre bundles} $P:= H \times_{\psi} L \rightarrow H/K$, where $\psi: K \rightarrow L$ is a Lie group homomorphism, in a natural way by setting
\begin{align}
L_h : P \rightarrow P, \text{ }[h',l] \mapsto [L_h h',l]. \label{lea}
\end{align}
Clearly, $L$ acts fibre transitively on $P$ and factorises to the transitive $H$-action $l$ on $M$. Main examples for this construction in this text are $TM=H\times_{Ad^{H/K}} \mathfrak{n}$ and the (real or complex) \text{spinor bundle} \[ S^g = \mathcal{Q}^g \times_{Spin^+} \Delta_{\mathfrak{n}} \cong H \times_{\widetilde{Ad}^{H/K}} \Delta_{\mathfrak{n}}.\]This in turn induces a natural $H-$left action on connections over homogeneous $G-$principal bundles $P \rightarrow H/K$, and we say that a connection $A \in \Omega^1(P,\,\mathfrak{g})$ is \textit{left-invariant} if it is invariant under this $H-$action, i.e.
\begin{align*}
L_h^* A = A \text{ }\forall h \in H.
\end{align*}

\begin{remark} \label{rr1}
In the case of the tangent bundle $TM$ it is straightforward to check that under the isomorphism (\ref{tb}) the canonical left action on $TM$ defined by (\ref{lea}) is the differential of the left action on $M$ i.e.
$L_h = dl_h$.
\end{remark}

\section{Holonomy of left-invariant spinor connections} \label{pura}
Let $(M,g)$ be a pseudo-Riemannian reductive homogeneous spin manifold of signature $(p,q)$ with fixed decomposition $M=H/K$, $\mathfrak{h} = \mathfrak{k} \oplus \mathfrak{n}$ and $Ad^{H/K}(k)$-invariant inner product $\langle \cdot, \cdot \rangle_{\mathfrak{n}}$ on $\mathfrak{n}$. Let $\nabla :\Gamma(S^g) \rightarrow \Gamma(T^*M \otimes S^g)$ denote the spinor covariant derivative, induced by the connection $\omega^{sp} \in \Omega^1 \left (\mathcal{Q}^g,\mathfrak{spin}(p,q)\right)$, which in turn is given as the lift of the Levi Civita connection $\omega^g$, i.e. by the commutative diagram
\begin{align*}
\begin{xy}
  \xymatrix{
      T\mathcal{Q}^g_+ \ar[r]^{\omega^{sp}} \ar[d]_{df^g}    &  \mathfrak{spin}(p,q) \ar[d]^{\lambda_*}  \\
      T\Pe^g \ar[r]_{\omega^g}            &   \mathfrak{so}(p,q).  
  }
\end{xy}
\end{align*}

Our goal is the determination of spinor fields which are parallel wrt. the \textit{modified covariant derivative}
\begin{align*}
D&=\nabla + \Omega: \Gamma(S^g) \rightarrow \Gamma(T^*M \otimes S^g),\\
D_X \ph &= \nabla_X \ph + \Omega(X) \cdot \ph \text{ for }X \in \mathfrak{X}(M), \ph \in \Gamma(S^g),
\end{align*}
where $\Omega: TM \rightarrow CL(TM,g) \cong \Lambda^*(TM)$ is a vector bundle homomorphism which is \textit{left-invariant}, i.e. $L_h^* \Omega = \Omega$, or in more detail
\begin{align}
l^*_{h^{-1}} \left(\Omega \left(dl_h(X)\right) \right) = \Omega(X) \text{ for all }X \in TM, h \in H. \label{dufff}
\end{align}

The main examples we will be dealing with are
\begin{enumerate}
\item $\Omega (X)= \lambda \cdot X^{\flat}$ for $\lambda \in \C$, which leads to the equation for geometric Killing spinors (cf. \cite{bfkg,kath}). In this case (\ref{dufff}) is vacuous.
\item $(M,g)$ is 11-dimensional Lorentzian and $\Omega(X) = c_1 X^{\flat} \wedge F + c_2 X \invneg F$, where $F$ is a closed 4-form and $c_{1,2}$ are real constants whose values depend on the use of a mostly + or mostly - metric. (\ref{dufff}) is satisfied if $F$ is $H-$invariant, i.e. $l_h^*F = F$ for $h \in H$ which on a symmetric space already implies that $F$ is actually parallel. This term arises within the context of 11-dimensional supergravity and will be studied in more detail in the next section.
\end{enumerate}

The problem of determining parallel sections wrt. $D$ can be addressed as follows: First, we show that $D$ is induced by a $H-$\textit{invariant} connection on the homogeneous $Cl^*(\mathfrak{n},\langle \cdot, \cdot \rangle_{\mathfrak{n}})$-bundle $H \times_K Cl^*(\mathfrak{n},\langle \cdot, \cdot \rangle_{\mathfrak{n}})$, where $K \rightarrow Cl^*(\mathfrak{n},\langle \cdot, \cdot \rangle_{\mathfrak{n}})$ acts by trivial extension of $\widetilde{Ad}^{H/K}:K \rightarrow Spin^+(\mathfrak{n}) \subset Cl^*(\mathfrak{n},\langle \cdot, \cdot \rangle_{\mathfrak{n}})$ and $Cl^*$ denotes the Clifford group. Second, we use results from \cite{kn1,kn2} which allow us to compute the holonomy algebra and curvature of invariant connections in a purely algebraic way.\\
\newline
To this end, we introduce some notation: The \textit{canonical 1-form} $\theta \in \Omega^1(\mathcal{P}^g,\R^{n})$ on $\mathcal{P}^g$ is given by\footnote{Given a $G-$principal bundle $P \rightarrow M$, an associated fibre bundle $P \times_G L \rightarrow M$ and $u \in P_x$, we denote by $[u]:L \rightarrow P_x$ the fibre isomorphism given by $l \mapsto [u,l]$.}
\begin{align*}
\theta_u(V):= [u]^{-1} d\pi_{\mathcal{P}^g \rightarrow M} (V).
\end{align*}
Moreover, we view $\Omega$ also as a section in the associated bundle $\mathcal{P}^g \times_{SO^+(p,q)} \text{End}(\R^n, Cl(p,q))$. Turning to spinors, we let $\widetilde{\theta}:= \left(f^g \right)^* \theta \in \Omega^1(\mathcal{Q}^g,\R^n)$ and $\rho: Cl(p,q) \rightarrow \text{End }(\Delta_{p,q})$ denote Clifford multiplication. By means of some pseudo-orthonormal basis we will for the subsequent calculation sometimes identify $(\mathfrak{n}, \langle \cdot, \cdot \rangle_{\mathfrak{n}}) \cong (\R^n, \langle \cdot, \cdot, \rangle_{p,q})$.\\
\newline
As a direct consequence of the various definitions, we obtain the following local expression for $D$: Let $s: U \subset M \rightarrow \mathcal{P}^g$ be a local section with lift $\widehat{s}:U \rightarrow \mathcal{Q}^g$. Let $X \in \mathfrak{X}(M)$ and $\ph \in \Gamma(S^g)$. Then on $U$ we may write $\ph = \left[ \widehat{s}, v\right]$ and $X = [s, t]$ for functions $v: U \rightarrow \Delta_{p,q}$ and $t:U \rightarrow \R^{p,q}$. It follows that
\begin{align}
\left(D_X \ph\right)_{|U} = \left[ \widehat{s}, t(v) + \rho_{*} \left(\omega^{sp}(d\widehat{s}(X))+[\pi_{T\mathcal{P}^g \rightarrow \mathcal{P}^g} (df^g d\widehat{s}(X)) ]^{-1} \left( \Omega \right) \left({\widetilde \theta}(d\widehat{s}(X)) \right) \right) (v) \right]. \label{lofo}
\end{align}
This motivates us to introduce the 1-form $\widetilde{\omega}:T \mathcal{Q}^g \rightarrow CL(p,q) \cong \Lambda^*\R^n$, given by
\begin{align}
\widetilde{\omega}(\widetilde{V}):= \left([\pi_{T\mathcal{P}^g \rightarrow \mathcal{P}^g} (df^g(\widetilde{V})) ]^{-1} \Omega \right) \left(\widetilde{\theta} \left (\widetilde{V}\right)\right). 
\end{align}
Using $R_A^* \theta (V) = A^{-1} (\theta(V)$ for $A \in SO^+(p,q)$ (cf. \cite{cs}), it is straightforward to calculate that for every $g \in Spin^+(p,q)$ we have
\begin{align}
R_g^* \widetilde{\omega} = Ad\left(\lambda(g)^{-1} \right) \circ \widetilde{\omega}. \label{ade}
\end{align}
$\widetilde{\omega}$ is invariant under the left-action of $H$, i.e. $L_h^* \widetilde{\omega} = \widetilde{\omega}$ for $h \in H$, which can be seen as follows:
Let $V_p \in T_p\mathcal{P}^g.$ Then 
\begin{align*}
L_h^* \theta \left(V_p \right) = \theta \left(dL_h V_p \right) = \left[ L_h(p) \right] d\pi_{\mathcal{P}^g \rightarrow M} dL_h V_p = \left[ L_h(p) \right] dl_h d\pi_{\mathcal{P}^g \rightarrow M}V_p
\end{align*}
Suppose that $\theta \left(V_p \right)=t$, i.e. $d\pi V_p = \sum_i t_i s_i$, where $p=(s_1,...,s_n) \in \mathcal{P}^g$. Then $dl_h d\pi V_p = \sum_i t_i dl_h(s_i)$ and the coefficients of this vector wrt. the frame $L_h(p) = (dl_h(s_1),...,dl_h(s_n))$ (cf. Remark \ref{rr1}) are clearly again given by $t$, i.e.  $L_h^* \theta = \theta$. It follows that 
\begin{align*}
L_h^* \widetilde{\omega} \left(\widetilde{V}_q \right) &= \left(\left[L_h (f^g(q)) \right]^{-1}\Omega \right)\left(\widetilde{\theta} \left(\widetilde{V}_q\right) \right)= \widetilde{\omega} \left(\widetilde{V}_q \right),
\end{align*}
where the last equality is equivalent to the requirement (\ref{dufff}).\\
\newline
We now extend $\mathcal{Q}^g$ to $\mathcal{Q}^{g,ext}:= \mathcal{Q}^g \times_{Spin^+(p,q)} Cl^*(p,q) \cong H \times_{\widetilde{Ad}^{H/K}} Cl^*(\mathfrak{n},\langle \cdot, \cdot \rangle_{\mathfrak{n}})$. Then the Levi Civita spin connection $\omega^{sp}$ naturally extends to a connection $\omega^{sp,ext}:T\mathcal{Q}^{g,ext} \rightarrow Cl(p,q)$, uniquely determined by the requirement
\begin{align}
\text{ker }\omega^{sp}_q = \text{ker }\omega^{sp,ext}_q \text{ for }q=[q,e] \in \mathcal{Q}^g \subset \mathcal{Q}^{g,ext}.
\end{align}
$\omega^{sp,ext}$ is left-invariant under the $H-$action as this holds for the Levi Civita connection $\omega^g$ which in turn is just a direct consequence of the behaviour of $\nabla^g$ under isometries. The $H-$invariant 1-form $\widetilde{\omega}$ can be naturally extended to $T\mathcal{Q}^{g,ext}$ as follows: Let $p = q \cdot g \in \mathcal{Q}^{g,ext}$, where $q \in \mathcal{Q}^g$ and $g \in Cl^*(p,q)$. We let $\widetilde{\omega}^{ext}$ be zero on vertical tangent vectors. For $Y \in \text{ker }\omega^{sp,ext}_q$ we find $Y' \in \text{ker }\omega^{sp}_p$ such that $dR_g Y' = Y$ and set $\widetilde{\omega}^{ext}(Y):= Ad(g^{-1}) \widetilde{\omega}(Y')$. It is straightforward to check that this is well-defined, i.e. independent of the choice of $g$ and $p$ and that $\widetilde{\omega}^{ext}$ is a $Ad-$equivariant (as follows from \ref{ade}) and horizontal (as follows from the definition of $\theta$)1-form on $\mathcal{Q}^{g,ext}$. 
Moreover, left invariance carries over to this natural extension.\\

In summary, $\omega^{ext}:= \omega^{sp,ext} + \widetilde{\omega}^{ext} : T\mathcal{Q}^{g,ext} \rightarrow Cl(p,q)$ is a $H-$invariant connection on the $H-$homogeneous principal bundle $H \times_{\widetilde{Ad}^{H/K}} Cl^*(\mathfrak{n},\langle \cdot, \cdot \rangle_{\mathfrak{n}}) \rightarrow H/K$. Clearly, the naturally associated covariant derivative on the bundle $S^g \cong \mathcal{Q}^{g,ext} \times_{\rho} \Delta_{p,q}$, where $\rho : Cl(p,q) \rightarrow End(\Delta_{p,q})$ is the standard representation\footnote{It might be unique up to equivalence or one has to require that the volume element is mapped to $\pm 1$}, is just $D$, as can be seen from the local formula (\ref{lofo}). Consequently, the parallel spinors of $D$ are equivalently encoded in the trivial sub-representations of $\rho$ restricted to the holonomy group $Hol(\omega^{ext}) \subset Cl^*(p,q)$. \newline

According to \cite{kn1,kn2} the holonomy algebra $\mathfrak{hol}(\omega^{ext})$ can be determined as follows\footnote{We use the notation from \cite{hammerl1,hammerl2} where this construction is reviewed. Moreover, this reference presents some examples and shows how the procedure can be applied to certain Cartan geometries which allows the determination of the conformal holonomy algebra of conformal structures over homogeneous spaces.}: In general, let $\gamma$ be a connection on a $P-$principal bundle $H \times_K P \rightarrow H/K$ \footnote{In fact, for every $P-$principal bundle $Q \rightarrow H/K$ on which $H$ acts fibre-transitively there exists a morphism $\psi : K \rightarrow P$ such that $Q = H \times_K P$, cf. \cite{hammerl1}} which is invariant under the $H-$action. We can associate to $\gamma$ a linear map\footnote{The precise correspondence between invariant connections and linear maps of this form is explained in \cite{kn1,kn2}.} $\alpha: \mathfrak{h} \rightarrow \mathfrak{p}$, given by $\alpha(X):=\widehat{\gamma}(e,e,X,0)$, where $\widehat{\gamma}$ denotes the extension of $\gamma$ to a connection on $H \times P$ and we trivialise $T(H \times P) =H \times P\times \mathfrak{h} \times \mathfrak{p}$. It holds that $\widehat{\gamma} \left(\frac{d}{dt}_{|t=0}(h(t),p(t))\right) = \gamma \left( \frac{d}{dt}_{|t=0} \left[ h(t),p(t) \right] \right)$. One then introduces the curvature map $\kappa \in \Lambda^2 \mathfrak{n}^* \otimes \mathfrak{p}$, which measures the failure of $\alpha$ to be a Lie algebra homomorphism, i.e.
\begin{align*}
\kappa(X_1,X_2) = \left[ \alpha(X_1),\alpha(X_2)\right]_{\mathfrak{p}} - \alpha \left( [X_1,X_2]_{\mathfrak{h}}\right).
\end{align*}
In this notation, the holonomy algebra $\mathfrak{hol}(\alpha):=\mathfrak{hol}(\gamma) \subset \mathfrak{p}$ of $\gamma$ can be calculated as follows: Let $\widehat{Im}(\kappa):= \text{span}(Im (\kappa)) \subset \mathfrak{p}$. $\mathfrak{hol}(\alpha)$ is the $\mathfrak{h}$-module generated by $\widehat{Im}(\kappa)$, i.e.
\begin{align*}
\mathfrak{hol}(\alpha) = \widehat{Im}(\kappa) + \left[\alpha(\mathfrak{h}),\widehat{Im}(\kappa)\right]+\left[\alpha(\mathfrak{h}),\left[\alpha(\mathfrak{h}),\widehat{Im}(\kappa)\right]\right]+...
\end{align*}
If actually $(M=H/K,g)$ is a symmetric space this simplifies to
\begin{align}
\mathfrak{hol}(\alpha) = \widehat{Im}(\kappa) + \left[\alpha(\mathfrak{n}),\widehat{Im}(\kappa)\right]+\left[\alpha(\mathfrak{n}),\left[\alpha(\mathfrak{n}),\widehat{Im}(\kappa)\right]\right]+... \label{holfo}
\end{align}

\begin{remark}
Note that (\ref{holfo}) generalises a well-known formula for the holonomy of the Levi Civita connection on symmetric spaces where $\alpha(\mathfrak{n}) = 0$, cf. \cite{eft}.
\end{remark}

Let us apply the computation of $\mathfrak{hol}$ to our original setting, i.e. $P=Cl^*(\mathfrak{n},\langle \cdot, \cdot \rangle_{\mathfrak{n}})$. The map $\alpha : \mathfrak{h} \rightarrow Cl(\mathfrak{n},\langle \cdot, \cdot \rangle_{\mathfrak{n}})$ decomposes into $\alpha=\alpha^{\mathfrak{k}}  + \alpha^{\mathfrak{n}}$ according to the decomposition $D = \nabla + \Omega$\footnote{We will justify this notation in a moment by checking that $\alpha^{\mathfrak{k}}$ lives only on $\mathfrak{k}$ and $\alpha^{\mathfrak{n}}$ only on $\mathfrak{n}$}. The map $\alpha^{\mathfrak{k}}$ which describes the Levi Civita spin connection has already been computed in \cite{hammerl2} for the general reductive homogeneous case\footnote{The computation presented there is for the Levi Civita connection only. However, passing to the induced spin connection is straightforward.}. In the following, we only need the result for $(H/K,g)$ being symmetric, and in this case $\alpha^{\mathfrak{k}}$ is simply the trivial extension of $\widetilde{ad}= \lambda_*^{-1} \circ ad :\mathfrak{k} \rightarrow \mathfrak{spin}(\mathfrak{n},\langle \cdot, \cdot \rangle_{\mathfrak{n}}) \subset Cl(\mathfrak{n},\langle \cdot, \cdot \rangle_{\mathfrak{n}})$.  In order to calculate the map $\alpha^{\mathfrak{n}}$ which encodes $\Omega$, we proceed as follows: Let $X \in \mathfrak{h}$ be generated by a curve $h(t)$ running in $H$ with $h(0)=e$. By the above definition of $\alpha$ we have
\begin{align*}
\alpha^{\mathfrak{n}}(X) &= \widetilde{\omega}( \frac{d}{dt}_{|t=0} \underbrace{\left[h(t),e \right]}_{ \in \mathcal{Q}^g = H \times_{\widetilde{Ad}} Spin^+(\mathfrak{n})} ) \\
&= ( [ \pi_{T \mathcal{P}^g \rightarrow \mathcal{P}^g} ( \frac{d}{dt}_{|t=0} {[h(t),e ]}) ]^{-1} \Omega ) (\theta (\frac{d}{dt}_{|t=0} [h(t),e ]) ) \\
&= ([ \underbrace{[e,e]}_{\in H \times_{Ad} SO^+(\mathfrak{n})}]^{-1}\Omega) ([[e,e]]^{-1} \underbrace{d\pi_{\mathcal{P}^g \rightarrow M} \frac{d}{dt}_{|t=0} [h(t),e ]}_{=\frac{d}{dt}_{|t=0}h(t)K} )\\
& = ([[e,e]]^{-1}\Omega) (\pi_{\mathfrak{h} \rightarrow \mathfrak{n}} X) \\
& = \Omega_{eK} ( \pi_{\mathfrak{h} \rightarrow \mathfrak{n}} X ) \in Cl(\mathfrak{n},\langle \cdot, \cdot \rangle_{\mathfrak{n}}) \cong \Lambda^* \mathfrak{n},
\end{align*}
i.e. $\alpha^{\mathfrak{n}}$ is the evaluation of $\Omega$ at the origin $eK$. The previous calculations reveal that the calculation of $\mathfrak{hol}(D)$ can be achieved by a purely algebraic algorithm.
Before we proceed, we make a technical remark: 

\begin{remark} \label{tr}
The target space of the maps $\alpha$ and $\kappa$ in our case is $Cl(\mathfrak{n},\langle \cdot, \cdot \rangle_{\mathfrak{n}}) \cong \Lambda^*(\mathfrak{n})$. Whence, if we compute the bracket $[\alpha(X_1),\alpha(X_2)]_{Cl(\mathfrak{n},\langle \cdot, \cdot \rangle_{\mathfrak{n}})} = \alpha(X_1) \cdot \alpha(X_2) - \alpha(X_2) \cdot \alpha(X_1)$, we would have to use the Clifford relations in the algebra $Cl(\mathfrak{n},\langle \cdot, \cdot \rangle_{\mathfrak{n}})$. However, we prefer working with the standard Clifford algebra $Cl(p,q)$ instead. This can be achieved by fixing an orientation preserving isometry $\eta: (\R^{p,q}, \langle \cdot, \cdot \rangle_{p,q}) \rightarrow (\mathfrak{n}, \langle \cdot, \cdot \rangle_{\mathfrak{n}})$, or equivalently, an oriented orthonormal basis of $\mathfrak{n}$,  considering the map $\alpha_{\eta}(X):= \eta^*(\alpha(X)) \in \Lambda^*\R^{p,q} \cong Cl(p,q)$, and carrying out all computations with $\alpha_{\eta}$ and $\kappa_{\eta}$. The curvature is in this picture given by
\begin{align*}
\kappa_{\eta} (X_1,X_2) = \left[\alpha^{\mathfrak{n}}_{\eta}(X_1),\alpha^{\mathfrak{n}}_{\eta}(X_2) \right]_{Cl(p,q)} - \lambda^{-1}_* \left(  \eta_{*}^{-1}\underbrace{ad [X_1,X_2]}_{\in \mathfrak{so}({\mathfrak{n}} )}\right) \in Cl(p,q).
\end{align*}
\end{remark}

We describe an example with non-trivial $\Omega$ which illustrates how the above procedure works in general. 

\begin{example}
We consider the standard sphere $S^n = SO^+(n+1)/SO^+(n)$ for which we have
\begin{align*}
\mathfrak{h}=\mathfrak{so}(n+1) = \left\{ \begin{pmatrix} 0 & -v^t \\ v & A \end{pmatrix} \mid A \in \mathfrak{so}(n) = \mathfrak{k}, v \in \mathfrak{n} \cong \R^n \right\}.
\end{align*} 
We want to show that $S^n$ admits geometric Killing spinors $\ph \in \Gamma(S^g)$ solving $\nabla_X \ph + \lambda X \cdot \ph = 0$, i.e. in this case $\Omega(X) = - \lambda \cdot X^{\flat}$. Fixing the standard inner product $\langle \cdot, \cdot \rangle_{st}$ on $\mathfrak{n}$ produces the sphere of Radius 1 with scalar curvature $n(n-1)$. We now let $c$ be a positive constant and fix the $Ad(K)$-invariant inner product $c \cdot \langle \cdot, \cdot \rangle_{st}$ yielding a sphere of scalar curvature $R=\frac{1}{c} n(n-1)$.\\
The adjoint action of $K$ on $\mathfrak{n}$ is simply the identity map. We further fix the isometry $\eta: \R^n \rightarrow \mathfrak{n}$, given by multiplication with $\frac{1}{\sqrt{c}}$. We then compute for $X_1,X_2 \in \mathfrak{n}$:
\begin{align*}
\kappa_{\eta}(X_1,X_2) &= \lambda^2 \cdot c \cdot [X_1,X_2]_{cl(n+1)} - \lambda_*^{-1} [X_1,X_2]_{\mathfrak{so}(n+1)} \\
& = \left( \lambda^2c - \frac{1}{4} \right) \cdot (X_1 \cdot X_2 - X_2 \cdot X_1)
\end{align*}
Thus, $\kappa \equiv 0$ iff $\lambda = \frac{1}{2 \sqrt{c}}$, i.e. $R=4n(n-1)\lambda^2$, meaning that the $S^n$ with scalar curvature $R$ admits a full space of Killing spinors to the Killing number $\pm \frac{1}{2} \cdot \sqrt{\frac{R}{n(n-1)}}$ which reproduces a well-known result.
\end{example}

\begin{remark}
Geometric Killing spinors on pseudo-Riemannian reductive homogeneous spaces and construction principles are further studied in \cite{kath}.
\end{remark}

\section{Application to symmetric M-theory backgrounds}
\label{symback}
Let $(M,g,F)$ be a classical $M-$theory background, i.e. $(M,g)$ is a 11-dimensional connected Lorentzian spin manifold equipped with a mostly minus-metric\footnote{The price we have to pay for this convention is that spheres have negative scalar curvature and $AdS$ spaces have positive scalar curvature} and $F$ is a closed $4-$form. The signature convention allows us to work with a real Clifford representation and real spinors $\Delta_{10,1}^{\R} \cong \R^{32}$ (cf. the Appendix for our conventions regarding Clifford algebras and spinors). $(M,g,F)$ satisfies the bosonic field equations

\begin{align}
d \ast F &= \frac{1}{2} F \wedge F,\label{fwedge} \\
Ric(X,Y) &= \frac{1}{2} \langle X \invneg F, Y \invneg F \rangle - \frac{1}{6} g(X,Y) |F|^2.
\end{align}

Setting the gravitino variation to zero in a purely bosonic background yields the Killing spinor equation
\begin{align}
D_X \ph = \nabla_X \ph + \underbrace{\frac{1}{6}(X \invneg F) \cdot \ph + \frac{1}{12} \left(X^{\flat} \wedge F\right) \cdot \ph}_{=:\Omega(X) \cdot \ph}. \label{sugra}
\end{align}
A background of 11-dimensional supergravity is called \textit{supersymmetric} iff it admits non-trivial solutions to (\ref{sugra}). \\

We shall now be concerned with \textit{symmetric} $M-$theory backgrounds $M=H/K$ and require $F$ to be $H-$invariant.  This implies that $F$ is parallel wrt. a torsion-free connection and so (\ref{fwedge}) reduces to the algebraic relation $F \wedge F = 0$.  These backgrounds have been studied in detail in \cite{jose}. However, it is yet unclear which of the classical backgrounds presented there admit solutions to (\ref{sugra}). We want to study this problem using the results from the previous section, i.e. we want to determine the symmetric $M-$theory backgrounds which preserve some supersymmetry.

\begin{remark}
In the following analysis, we ignore all backgrounds from  \cite{jose} which involve a $\mathbb{C}P^2-$ or $\mathbb{C}H^2$- factor. They are solutions to the field equations but $\mathbb{C}P^2$ does not admit a spin structure, whence it makes no sense to speak about supersymmetry in this case. The noncompact dual $\mathbb{C}H^2$ does not admit a homogeneous spin structure as its isotropy representation is equivalent to that of $\mathbb{C}P^2$. 
\end{remark}

\begin{example} \label{deb}
As a first example we show how the algebraic algorithm from section \ref{pura} reproduces the well-known fact that the \textit{Freund-Rubin background},
\begin{align}
(M,g) = (M_1,g_1) \times (M_2,g_2) = AdS_7(7R) \times S^4(-8R), F= \sqrt{6R}  \cdot vol S_4
\end{align} where $-R < 0$ is the constant scalar curvature of $(M,g)$, is maximally supersymmetric. In the following, subscript $1$ refers to the $AdS_7$-factor, subscript $2$ to the $S^4$-factor. We have $\mathfrak{n}=\mathfrak{n}_1 \oplus \mathfrak{n}_2$ etc.\\
$AdS_7 = SO^+(6,2)/SO^+(6,1)$ similar to the $S^n$ example discussed before. In particular, the adjoint action of $K_1$ on $\mathfrak{n}_1$ is again given by the identity. In order to obtain scalar curvature $7R$, one has to fix the invariant inner product $\frac{6}{R} \langle \cdot, \cdot \rangle_{6,1}$ on $\mathfrak{n}_1 \cong \R^{6,1}$, i.e. a isometry between $\mathfrak{n}_1$ and $\R^{6,1}$ is in this picture given by multiplication with $\sqrt{\frac{6}{R}}$.
Similarly, in order to obtain scalar curvature $-8R$ on $S^4$, we have to fix the inner product $\frac{3}{2R} \cdot \langle \cdot, \cdot \rangle_{0,4}$ on $\mathfrak{n}_2$. We let $X_1,X_2 \in \mathfrak{n}_1$ and $Y_1,Y_2 \in \mathfrak{n}_2$. Note that $\widetilde{ad}([X_1,X_2]) = \frac{1}{4} \left(X_1 \cdot X_2 - X_2 \cdot X_1 \right)$, whereas $\widetilde{ad}([Y_1,Y_2]) = -\frac{1}{4} \left(Y_1 \cdot Y_2 - Y_2 \cdot Y_1 \right)$. To see this, let $Z \in \{X,Y\}$ and note that $\widetilde{ad}([Z_1,Z_2])=\frac{1}{4} \left(Z_1 \cdot e_Z \cdot Z_2 \cdot e_Z - Z_2 \cdot e_Z \cdot Z_1 \cdot e_Z \right) = -\frac{1}{4}e_Z^2 \left( Z_1Z_2 -Z_2Z_1 \right)$, where $e_Z^2 = -1$ for $Z=X$ and $e_Z^2=+1$ for $Z=Y$. Moreover, using the above isometric identifications, we compute
\begin{align*}
\alpha^{\mathfrak{n}_1}(X_i) = \frac{1}{12} \sqrt{\frac{6}{R}} \sqrt{6R} X_i^{\flat} \wedge vol_{\R^4} = \frac{1}{2}X_i \cdot vol_{\R^4} = \frac{1}{2} vol_{\R^4} \cdot X_i, \\
\alpha^{\mathfrak{n}_2}(Y_i) = \frac{1}{6} \sqrt{\frac{3}{2R}} \sqrt{6R} Y_i \invneg vol_{\R^4} = - \frac{1}{2}Y_i \cdot vol_{\R^4} = \frac{1}{2} vol_{\R^4} \cdot Y_i. \\
\end{align*}
Together with $d vol_{\R^4}^2 = 1$ it follows that $\left[ \alpha^{\mathfrak{n}_1}(X_i),\alpha^{\mathfrak{n}_1}(X_j)\right] = \frac{1}{4}(X_iX_j-X_jX_i)$, $\left[ \alpha^{\mathfrak{n}_2}(Y_i),\alpha^{\mathfrak{n}_2}(Y_j)\right] = -\frac{1}{4}(Y_iY_j-Y_jY_i)$ and $\left[ \alpha^{\mathfrak{n}_1}(X_i),\alpha^{\mathfrak{n}_2}(Y_j)\right] = 0$ such that $\kappa \equiv 0$, and thus also $\mathfrak{hol}(\alpha) = 0$. A similar analysis can be carried out for the maximally supersymmetric background $AdS_4(8R) \times S^7(-7R)$ and $F=\sqrt{6R} vol_{AdS_4}$. 
\end{example}

Using the theory developed in section \ref{pura}, we describe how the algorithm which computes $\mathfrak{hol}(D) \subset \mathfrak{gl}(32,\R)$ \footnote{In fact, one can show that for every  background we have $\mathfrak{hol}(D) \subset \mathfrak{sl}(32,\R)$} works in practice:\\
\newline
For the symmetric space $(M=H/K,g)$, which will be a metric product of one indecomposable Lorentzian symmetric space and zero or more irreducible Riemannian symmetric spaces, suppose that the Lie algebras $\mathfrak{h}=\mathfrak{k} \oplus \mathfrak{n}$ with structure constants are given. First, determine the $Ad(K)-$invariant inner product on $\mathfrak{n}$ corresponding to $g$. Let 
\begin{align*}
T_1,...,T_n \text{ be a pos. oriented orthonormal basis of }\mathfrak{n},\\
L_1,...,L_m \text{ be a basis of }\mathfrak{k},
\end{align*}
such that for certain constants
\begin{align}
\left[ T_i,T_j \right] = \sum_k c^{ij}_k L_k \text{ and } \left[ L_k,T_i\right] = \sum_v d^{ki}_v T_v.
\end{align}
Further, suppose that wrt. the fixed orthonormal basis the 4-form $f = F_e$ is on $\R^{10,1}$ \footnote{In the notation of Remark \ref{tr}, $f = \eta^*\left(F_e \right)$} given as
\begin{align}
f = \sum_{\nu_1 <...<\nu_4} f_{\nu_1,...,\nu_{4}} e_{\nu_1}^{\flat} \wedge...\wedge e_{\nu_4}^{\flat},
\end{align}
where $e_i$ is the standard basis of $\R^{11}$ and $e_i^\flat(e_j) = \langle e_i,e_j \rangle \delta_{ij}$. Fix a concrete Clifford representation, i.e. matrices $\rho(e_i) \in GL(32,\R)$, and calculate for $1 \leq i \leq 11$ the Clifford products 
\begin{align}
\alpha_F(e_i):=\sum_{\nu_1 <...<\nu_4} f_{\nu_1,...,\nu_{4}} \left(-\frac{1}{24} e_i \cdot e_{\nu_1}\cdot...\cdot e_{\nu_4} + \frac{1}{8} e_{\nu_1}\cdot...\cdot e_{\nu_4} \cdot e_i\right) \in M(32,\R),
\end{align}
and then compute for $1 \leq i , j \leq 11$ the matrices
\begin{align}
\kappa(e_i,e_j) = -\frac{1}{2} \cdot \sum_k\sum_{l<v} c^{ij}_k d^{kl}_v e_l \cdot e_v + \left[\alpha_F(e_i),\alpha_F(e_j) \right] \in M(32,\R). \label{kako}
\end{align}
The linear span of the elements (\ref{kako}) gives the vector space $\widehat{Im}(\kappa) \subset M(32,\R)$. Fix a basis of this space and compute the commutator with all elements  $\alpha_F(e_i)$ leading to the space $[\widehat{Im}(\kappa),\alpha(\mathfrak{n})]$.  Iterate this, i.e. compute $[[\widehat{Im}(\kappa),\alpha(\mathfrak{n})],\alpha(\mathfrak{n})]$ etc. until the dimension becomes stable. Then by (\ref{holfo}) the holonomy algebra is found as a matrix subalgebra of $M(32,\R)$ and one can now compute all spinors annihilated by it. By the holonomy principle they correspond to $D-$parallel spinors (in the simply-connected case).

\begin{remark} 
Note that for a (simply-connected) symmetric M-theory background of the form $(M,g)=(H/K,g)$ such that\footnote{Every $X \in \mathfrak{h}$ generates a Killing vector field on $M$ by setting $X^*(hK):= \frac{d}{dt}_{|t=0} \text{exp}(tX) hK$. We assume that these are all Killing vector fields.} $\mathfrak{h} \cong Kill(M,g)$, the determination of the associated \textit{Killing superalgebra} (cf. \cite{jose2}) is also purely algebraic. The odd part is $\mathfrak{g}_1 = \{ v \in \Delta_{10,1}^{\R} \mid \mathfrak{hol} (D) \cdot v =0 \}$ and the even part is given by $\mathfrak{g}_0 = \mathfrak{h} = \mathfrak{k} \oplus \mathfrak{n}$, where we use the isomorphism to $Kill(M,g)$ given by
\begin{align}
Kill(M,g) \ni X \mapsto \left(\left(\nabla X \right)_{eK}, X(eK) \right),
\end{align}
which identifies $\mathfrak{h}$ with a subspace of $\mathfrak{so}(\mathfrak{n})$. The brackets can now be computed as follows: The even-even bracket is simply (minus) the bracket in $\mathfrak{h}$. The odd-even bracket is classically given as 
\begin{align*}
L_{X} \ph := \nabla_X \ph + \frac{1}{2} \underbrace{\left(\nabla(X) \right)}_{\in \mathfrak{so}(TM) \cong \Lambda^2(TM)} \cdot \ph = - \Omega(X) \cdot \ph + \frac{1}{2} {\left(\nabla(X) \right)} \cdot \ph
\end{align*}
Thus, under the above identifications this corresponds to
\begin{align}
\g_0 \otimes \g_1 \ni (\alpha,t) \otimes v \mapsto \left(\frac{1}{12}  t^{\flat} \wedge f + \frac{1}{6} t \invneg f + \frac{1}{2} \alpha \right) \cdot v \in \g_1,
\end{align}
where as usual $f = F_{eK} \in \Lambda^4 \R^{10,1}$. Using formulas from \cite{jose2}, we further conclude that under the above isomorphisms the symmetric odd-odd-bracket is given by
\begin{align*}
\g_1 \otimes \g_1 \ni v_1 \otimes v_2 \mapsto (\alpha_{v_1,v_2},t_{v_1,v_2}) \in \g_0,
\end{align*}
where the spinor bilinears are given by $\langle t_{v_1,v_2}, e_i \rangle = \langle v_1, e_i \cdot v_2 \rangle_{\Delta} \text{ and } \langle \alpha_{v_1,v_2} e_i^{\flat} \wedge e_j^{\flat} \rangle = -\frac{1}{3} \langle v_1, e_i \invneg e_j \invneg f \cdot v_2 \rangle_{\Delta} + \frac{1}{6} \langle v_1, e_i \wedge e_j \wedge f \cdot v_2 \rangle_{\Delta}$.
This determines the bracket on $\g$ completely and the structure of $\g$ can then be further analysed via its Levi decomposition.
\end{remark}

\section{Exclusion of backgrounds}
\label{exclude}
Our aim in this section is to show how the algorithm derived in the previous sections can be used to exclude a large class of symmetric $M-$theory backgrounds from preserving supersymmetry. This will follow from elementary algebraic observations. The remaining backgrounds will then be attacked computationally.  In the following, we always assume that the underlying symmetric space is given as a metric product of \textit{non-trivial} symmetric spaces
\begin{align}
(M,g)= \left(H/K,g \right) = ( H_1/K_1,g_1) \times...\times (H_k/K_k,g_k) \label{45}
\end{align}
with decompositions $\mathfrak{h}_j = \mathfrak{k}_j \oplus \mathfrak{n}_j$ for $j=1,...,k$. We always identify $\mathfrak{n}_j \cong \R^{\text{dim }\mathfrak{n}_j}$ by means of a positively-oriented orthonormal basis (cf. Remark \ref{tr})
\begin{remark}
Suppose we can show that $\mathfrak{spin}(\mathfrak{n}_1) \subset \mathfrak{hol}(\alpha) \subset Cl(10,1)$. This implies that $Spin^+(\mathfrak{n}_1) \times 1 \times....\times 1 \subset Hol(\alpha) \subset Cl^*(10,1)$. Now a general Lemma (cf. \cite{leiphd}) states that if $\rho$ is a representation of a group $G=G_1 \times G_2$ and $\rho_i$ are representations of $G_i$ such that $\rho \propto \rho_1 \times \rho_2$, then $\rho$ has a trivial sub-representation iff both $\rho_i$ has a trivial sub-representations. This can immediately be applied to our setting. More precisely, consider the group $G:= \left(Spin(r_1,s_1) \times Spin(r_2,s_2) \right) / \mathbb{Z}_2 \hookrightarrow Spin(r:=r_1+r_2, s:=s_1+s_2)$. Then we have two natural representations of $G$: The first one is given as the tensor product $\rho$ on the spinor representations $\Delta_{r_i,s_i}$, the second, $\Phi$ results from the restriction of the spinor representation $\Phi: Spin(r,s) \rightarrow End \left(\Delta_{r,s} \right)$ to $G$. One can show that $\rho \propto \Phi$.  A nonzero $Hol(\alpha)$-invariant spinor would lead to a nonzero spinor in $\Delta_{\mathfrak{n}_1}$ fixed by $Spin(\mathfrak{n}_1)$ which does not exist. Consequently, there are no $D$-parallel spinors in this situation.\\
\end{remark}

\begin{Lemma} \label{l1}
Given the decomposition (\ref{45}), assume in addition that $k \geq 3$ and that the invariant 4-form $F$ is the pullback of a 4-form on (w.l.o.g) $M_1$ which satisfies $F \cdot F = F^2 = \text{const}. \neq 0$ \footnote{In contrast to \cite{jose} we denote by $F^2$ the Clifford product $F \cdot F$ and \textit{not} the wedge product $F \wedge F$}. Then there is no $D-$parallel spinor on $(M,g)$.
\end{Lemma}

\textit{Proof: }
In what follows, we always let $f \in \Lambda^4 \mathfrak{n}^*_1$ denote the evaluation of $F$ at the origin $eK$. Further, let $X \in \mathfrak{n}_2$ and $Y \in \mathfrak{n}_3$. Then we have that $[X,Y]_{\mathfrak{h}}=0, X\cdot f = f \cdot X, Y \cdot f = f \cdot Y$, and therefore
\begin{align*} 
\kappa(X,Y) &= \frac{1}{12}\left[X \cdot f, Y \cdot f\right] = \frac{f^2}{6}X \cdot Y \in Cl(10,1), 
\end{align*}
i.e. $X \cdot Y \in \mathfrak{hol}(\alpha)$. As this space is a Lie algebra, it is easy to see that $X_1 \cdot X_2 \in \mathfrak{hol}(\alpha)$ for all $X_{i} \in \mathfrak{n}_1$, i.e. $\mathfrak{spin}(\mathfrak{n}_1) \subset \mathfrak{hol}(\alpha)$. The claim follows with the previous Remark.

\begin{remark}
The Freund Rubin backgrounds show that Lemma \ref{l1} does not hold in case $k \leq 2$.
Note that the conditions from Lemma \ref{l1} are satisfied if $k \geq 3$ and $F$ is a multiple of the volume form on $M_1$. 
\end{remark}

\begin{remark} \label{2}
Another background which is not covered by Lemma \ref{l1} is (4.7.2), $AdS_2 \times H^2 \times S^7$ with $F=f_0 \nu \wedge \sigma$ for some constant $f_0$ and $\nu$ and $\sigma$ being the area forms. Let $X \in \mathfrak{n}_{H^2}$. There is a nonzero constant $c$ such that $\Omega(X)=c \cdot X \cdot F = -c \cdot F \cdot X$. For $Y \in \mathfrak{n}_{AdS_2}$ we have $\Omega(Y)=c \cdot Y \cdot F = -c \cdot F \cdot Y$. By (\ref{holfo}) it follows that
\begin{align*}
\alpha(X,Y)&=[\Omega(X),\Omega(Y)] \\
&=c^2 (X \cdot F \cdot Y \cdot F - Y \cdot F \cdot X \cdot F) = -c^2 X \cdot Y \stackrel{!}{\in} \mathfrak{hol}(D).
\end{align*}
With the same conclusion as in Lemma \ref{l1}, there can be no holonomy-invariant spinor.
\end{remark}

\begin{remark} \label{3}
We exclude some backgrounds $(M,g)$ involving flat tori $T^n = \mathbb{R}^n / \mathbb{Z}^n$, from which $\mathfrak{k}_{T^n}=\{0\}$ follows, i.e. $\mathfrak{n}_{T^n}$ is the abelian Lie algebra $\R^n$. It is then a direct consequence of (\ref{holfo}) that every spinor $v$ annihilated by $\mathfrak{hol}(D)$ has to satisfy
\begin{align}
\kappa(X,Y) \cdot v = [\Omega(X),\Omega(Y)] \cdot v = 0 \text{ }\forall X,Y \in \mathfrak{n}_{T^n}. \label{str2}
\end{align}
Moreover, if we assume that there are linearly independent $X,Y \in \mathfrak{n}_{T^n}$ such that $X \cdot F = \pm F \cdot X$ and $Y \cdot F = \pm F \cdot Y$, we must even have that
\begin{align}
F\cdot F \cdot v = 0, \label{str}
\end{align}
i.e. $F \cdot F$ must have a kernel if considered as endomorphism acting on spinors. \\
To start with, consider $AdS_2 \times \mathbb{C}P^3 \times T^3$ with $F=f(\omega \pm \sqrt{3} \nu) \wedge d\theta^{12}$. Obviously, $\theta^i \cdot F = - F \cdot \theta^i$ for $i=1,2$. Thus (\ref{str}) must hold. However, $F \cdot F = -f^2 \cdot (\omega \cdot \omega \pm 2 \sqrt{3} \nu \cdot \omega +3)$. Fixing a concrete realisation of Clifford multiplication shows that ker $(F \cdot F) = 0$.\\
Next, we elaborate on (4.7.7): $AdS_2 \times S^3 \times S^3 \times T^3$ and $F=f(\sigma_3 \pm \sigma'_3) \wedge d\theta^3$. Also in this case (\ref{str}) must hold. However, 
\[F \cdot F = f^2 \cdot (-2),\]
as $\sigma_3 \cdot \sigma'_3 = - \sigma_3 \cdot \sigma'_3$. Thus, $F \cdot F$ has no kernel.\\
Similarly, by using (\ref{str2}) we also exclude the backgrounds (4.7.11). Finally, consider the background (4.7.15), i.e. $AdS_2 \times S^2 \times T^7$ and $F=f(\nu \pm \sigma_2)\wedge d\theta^{23}$. (\ref{str}) must hold, but $F \cdot F = 2f^2 \cdot \nu \cdot \sigma_2$ which obviously acts invertibly on spinors.
\end{remark}

\begin{remark} \label{rety}
We now show how further backgrounds can be excluded or how the region in the $F-$moduli space which allows supersymmetry can be restricted by an eigenvalue analysis of the endomorphism $F \cdot F \in Cl(10,1)$ by considering an example of backgrounds of the form $(4.7.12)$, i.e.
\begin{equation} \label{yu}
\begin{aligned} 
AdS_2 \times S^5 \times S^2 \times S^2, \\
f=F_e=f_0 \nu \wedge \sigma_1 + f_1 \nu \wedge \sigma_2 + f_2 \sigma_1 \wedge \sigma_2,
\end{aligned} 
\end{equation}
where $\nu$ is the area form on $AdS_2$, $\sigma_i$ are the area forms on the two spheres, which are also allowed to be tori or hyperbolic spaces. 
Inserting vectors which are tangent to the $S^5$-factor into $\kappa$ yields that every holonomy-invariant spinor $v$ has to satisfy
\begin{align}
F^2 \cdot v = c \cdot v \label{ft}
\end{align}
for some \textit{real} constant $c \in \R$. We use the fact that the $\sigma_i$ and $\nu$ commute with each other as well as $\sigma_i^2=-1$ and $\nu^2=+1$ to conclude with (\ref{yu}) that all eigenvalues of $F \cdot F$ are of the form
\begin{align}
(i\epsilon_0 f_0 + i\epsilon_1 f_1  \epsilon_2 f_2)^2,
\end{align}
where $\epsilon_{0,1} \in \{ \pm \}$. Setting the imaginary part to zero shows that either $f_2=0$ or $f_0^2 = f_1^2$. This directly excludes the backgrounds $AdS_2 \times SLAG_3 \times S^2 \times T^2$ and $AdS_2 \times S^5 \times S^2 \times T^2$, as our $f_i-$constraints are incompatible with the Einstein equations for the factors. Furthermore, we can exclude $AdS_2 \times SLAG_3 \times H^2 \times T^2$ and $AdS_2 \times S^5 \times H^2 \times T^2$ as follows: The condition $f_0^2=f_1^2$ is incompatible with the Einstein equations for the factors, thus we must have that $f_2=0$. The $T^2$-Einstein equation yields for this case that $2f_0^2 = \pm f_1^2$. That is, $F$ depends only on one overall prefactor and it is straightforward to calculate that $Im\text{ }\kappa$ does not annihilate any nonzero spinor in this case.
\end{remark}

\begin{remark} \label{ree}
The analysis from Remark \ref{rety} carries over to other possible backgrounds such as (4.7.14) and (4.7.16). In each of these cases, inserting vectors with $x \invneg f = 0$ into $\kappa$ yields eigenvalue equations of the form
\begin{align}
F^2 \cdot v = c \cdot v, \label{gfg}
\end{align} 
where $c \in \R \backslash \{0 \}$ is a constant which can be determined from the Einstein equations, i.e. $v$ is an eigenspinor to a nonzero real eigenvalue of $F^2$ which is known in terms of the $f_i$. The eigenvalues of $F^2$ can be easily calculated as done above since the action of the area forms on the spinor module is well known. Comparing with (\ref{gfg}) leads to additional constraints on the $f_i$ required for supersymmetry.
\end{remark}

\begin{remark} \label{1}
Next, we exclude backgrounds involving the K{\"a}hler form of a factor in $F$.
To start with, consider backgrounds of the form (\ref{45}) with $k \geq 3$ and $F$ being proportional to the square of the K{\"a}hler form, i.e.  $F= \omega \wedge \omega$ on the factor $M^1$. In this case a calculation which runs through the same lines as the proof of Lemma \ref{l1} reveals that every spinor fixed by the holonomy representation has to satisfy $F \cdot F \cdot v = 0$. However, the action of $\omega$ on the (complex) spinor module is well known (cf. \cite{kir}) and using these results it is straightforward to see that $F$ considered as endomorphism on the spinor module has no zero eigenvalue if $M_1$ is 6-dimensional. This excludes the backgrounds
\begin{align*}
AdS_{3,2} \times \mathbb{C}P^3 \times H^{2,3}, AdS_{3,2} \times G^+_{\R}(2,5) \times H^{2,3} (4.6.1.), (4.7.5).
\end{align*} 
Moreover, the backgrounds (4.4.1.), i.e. $AdS_5 \times \mathbb{C}P^3$ as well as $AdS_5 \times G_{\mathbb{R}}^+(2,5)$, where $F$ is a constant multiple of $\omega \wedge \omega$ can be excluded as follows:\\
Inserting a $AdS_3-$direction $X \in \mathfrak{n}_{AdS_3}$ and a $\mathbb{C}P^3-$direction $Y \in \mathfrak{n}_{AdS_3}$ into \ref{holfo} shows that every $\mathfrak{hol}(D)$-invariant spinor $v$ has to satisfy
\begin{align*}
[\Omega(X),\Omega(Y)] \cdot v =0.
\end{align*}
Fixing a concrete realisation of Clifford multiplication shows that this algebraic equation is satisfied only for $v=0$.
\end{remark}

\begin{remark} \label{s1}
We exclude the background (4.7.16), i.e. $M_0 \times M_1 \times...\times M_4 \times S^1:=AdS_2 \times S^2 \times S^2 \times S^2 \times S^2 \times S^1$, where $F=\sum_{i=1}^4f_i \nu \wedge \sigma_i + \sum_{i<j} f_{ij} \sigma_i \wedge \sigma_j$ for the area form $\nu$ of $AdS_2$ and $\sigma_i$ are the area forms of the respective spheres. For each $i=0,...,4$, we split $F=F^i_1+F^i_2$, where $F^i_1$ contains precisely those summands involving $vol_{M_i}$. Inserting nonzero vectors tangent to $S^1$ and $M_i$ into $\kappa$, yields that every $\mathfrak{hol}$-invariant spinor has to satisfy
\begin{align}
\left(2 \cdot F_1^i \cdot F_2^i +(F_2^i)^2\right) \cdot v \stackrel{!}{=}0. \label{trump}
\end{align}
To analyse this further, we choose bases of $\Delta_{1,1}$ and $\Delta_2$ in which $\nu$ and $\sigma_i$ become diagonal (with eigenvalues $\pm 1$ and $\pm i$, respectively). By forming tensor products, we obtain a basis of $\Delta_{10,1}$ in which 
\begin{align*}
2 \cdot F_1^i \cdot F_2^i +(F_2^i)^2 = \text{diag}(\mu^i_1,....,\mu^i_{32})
\end{align*}
for certain $\mu^i_k \in \mathbb{C}$. (\ref{trump}) translates into the existence of a fixed $k \in \{1,....,32\}$ such that $\mu^i_k = 0$ for all $i=0,...,5$. The $\mu^i_k$ are straightforward to compute. For a particular ordered basis, one finds that $\mu^l_1=0$ for $l=0,...,4$ is equivalent to
\begin{equation} \label{ev}
\begin{aligned}
0 &= \sum_{1 \leq i < j \leq 4} f_{ij}, \\
0 &= ( \sum_{j \neq i} f_j ) \cdot (2f_i +  \sum_{j \neq i} f_j ) - (\sum_{\substack{ 1 \leq k < l \leq 4, \\ k,l \neq i}} f_{kl})^2\text{ for }i=1,...,4, \\
0 &= f_i \cdot \sum_{\substack{ 1 \leq k < l \leq 4, \\ k,l \neq i}} f_{kl} \text{ for }i=1,...,4.
\end{aligned}
\end{equation}
All other equations $\mu^l_{k>1}=0$ are obtained from (\ref{ev}) by some sign changes, the resulting analysis is completely analogous to the case presented here: Imposing additionally the algebraic relations resulting from the Einstein equation $F \wedge F=0$, it is easy to verify that (\ref{ev}) has only solutions if $f_i=0$ for all $i$. However, as furthermore by the $S^1$-Einstein equations $\sum_{i}f_i^2 = \sum_{k<l}f_{kl}^2$, this implies that $F \equiv 0$. Note that these conclusions do not depend on the Einstein equations of the 2-dimensional factors. Whence the Remark also applies to the backgrounds in (4.7.16) which are obtained by replacing an $S^2$-factor with $H^2$ or $T^2$.
\end{remark}

We summarise the backgrounds excluded by the previous remarks and observations as follows:
\begin{longtable}[h]{|p{1cm}|p{4.5cm}|p{4.5cm}|p{2cm}|} \hline
Number (\cite{jose}) & Background & Invariant form $F$  & Reason for exclusion \\  \hline\hline
4.2 & $AdS_7 \times S^2 \times S^2$ & $f\cdot \omega_4$ & Remark \ref{1} \\ \hline
4.4.1 & $AdS_5 \times \mathbb{C}P^3,$ \newline $AdS_5 \times G_{\mathbb{R}}^+(2,5)$ & $\frac{1}{2}f \cdot \omega^2$ & Remark \ref{1} \\ \hline
4.4.2 & $AdS_5 \times H^2 \times S^4$ & $f \cdot \omega_4$ & Lemma \ref{l1} \\ \hline
4.5.1 & $AdS_4 \times S^5 \times S^2$, \newline $AdS_4 \times SLAG_3 \times S^2$ & $f \cdot \nu$ & Lemma \ref{l1} \\ \hline
4.5.2 & $AdS_4 \times S^4 \times S^3$, \newline $AdS_4 \times S^2 \times S^2 \times S^3$  & $f \cdot \nu$ & Lemma \ref{l1} \\ \hline
4.5.2 & $AdS_4 \times S^4 \times H^3$, \newline $AdS_4 \times S^2 \times S^2 \times H^3$ & $f \cdot \sigma$ & Lemma \ref{l1} \\ \hline
4.6.1 & $AdS_3 \times \mathbb{C}P^3 \times H^2$, \newline $AdS_3 \times G_{\mathbb{R}}^+(2,5) \times H^2$ & $\frac{1}{2}f \cdot \omega^2$ & Remark \ref{1} \\ \hline
4.6.2 & $AdS_3 \times S^4 \times H^4$ & $f \cdot \nu_{comp}$ & Lemma \ref{l1} \\ \hline 
4.6.2 & $AdS_3 \times S^4 \times H^2 \times H^2$, \newline $AdS_3 \times S^2 \times S^2 \times H^4$ & $f \cdot \nu_{comp}$ & Lemma \ref{l1} \\ \hline 
4.7.1 & $AdS_2 \times SLAG_4$ & $f \cdot \Omega$ & direct calculation \\ \hline
4.7.2 & $AdS_2 \times H^2 \times S^7$ & $f \nu \wedge \sigma$ & Remark \ref{2} \\ \hline
4.7.3 & $AdS_2 \times H^5 \times S^4$, \newline $AdS_2 \times (SL(3,\R)/SO(3)) \times S^4$ & $f \cdot \sigma_4$ & Lemma \ref{l1} \\ \hline
4.7.4 & $AdS_2 \times G_{\C}(2,4) \times S^1$ & $f \cdot(\sqrt{\frac{3}{2}} \nu \wedge \omega \pm (\Omega^{(1)} - \Omega^{(2)}))$ & direct calculation \\ \hline
4.7.5 & $AdS_2 \times H^3 \times \mathbb{C}P^3$ \newline $AdS_2 \times H^3 \times G_{\mathbb{R}}^+(2,5)$ & $\frac{1}{2}f \cdot \omega^2$ & Remark \ref{1} \\ \hline
4.7.6 & $AdS_2 \times \mathbb{C}P^3 \times T^3$, \newline $AdS_2 \times G_{\mathbb{R}}^+(2,5) \times H^3$ & $f(\omega \pm \sqrt{3} \nu) \wedge d\theta^{12}$ & Remark \ref{3}\\ \hline
4.7.7 & $AdS_2 \times S^3 \times S^3 \times T^3$ & $f(\sigma_3 \pm \sigma'_3) \wedge d\theta^3$ & Remark \ref{3} \\ \hline
4.7.8 & $AdS_2 \times S^4 \times S^3 \times H^2$ & $f \cdot \nu \wedge \sigma_2$ & Lemma \ref{l1} \\ \hline
4.7.8 & $AdS_2 \times S^4 \times H^3 \times H^2$ & $f \cdot \sigma_4$ & Lemma \ref{l1} \\ \hline
4.7.11 & $AdS_2 \times SLAG_3 \times T^4$, \newline $AdS_2 \times S^5 \times T^4$ & $f \cdot \left(d\theta^{1234} \pm \sqrt{2} \nu \wedge (d\theta^{12} \pm d\theta^{34}) \right)$ & Remark \ref{3} \\ \hline
4.7.12 \newline ($2^{nd}$ part) & $AdS_2 \times SLAG_3 \times S^2 \times T^2$, \newline $AdS_2 \times SLAG_3 \times H^2 \times T^2$, \newline $AdS_2 \times S^5 \times S^2 \times T^2$, \newline $AdS_2 \times S^5 \times H^2 \times T^2$ & $f_0 \nu \wedge \sigma_1 + f_1 \nu \wedge \sigma_2 + f_2 \sigma_1 \wedge \sigma_2$ & Remark \ref{rety} \\ \hline
4.7.16 & $AdS_2 \times S^2 \times S^2 \times S^2 \times S^2 \times S^1$, \newline  $AdS_2 \times H^2 \times S^2 \times S^2 \times S^2 \times S^1$, \newline  $AdS_2 \times H^2 \times H^2 \times S^2 \times S^2 \times S^1$ & $\sum_{i=1}^4f_i \nu \wedge \sigma_i + \sum_{i<j} f_{ij} \sigma_i \wedge \sigma_j$ & Remark \ref{s1} \\ \hline
\end{longtable}

\begin{remark}
After this analysis, there are a number of backgrounds for which the existence of supersymmetries is still undecided.  In all these cases, the 4-form $F$ depends on various parameters which are subject to additional algebraic equations or inequalities. The $F-$moduli space needed for supersymmetry can be reduced further as explained in Remark \ref{ree}. 
The resulting equations for the parameters $f_i$ appearing in $F$ are much more involved and so in order to complete our analysis we turn to the computer.
We have directly computed concrete realisations of $\kappa$ for each of these backgrounds\footnote{The source code for this software is available upon request.}.  Then, following the spirit of the previous analysis, we choose a convenient pair $X,Y \in \mathfrak{n}$ and compute the eigenvalues of $\kappa(X,Y)$.
\begin{itemize}
\item $AdS_5 \times S^2 \times S^2 \times S^2 / H^2$ (4.4.3): We take three generators; one constructed from two vectors on the $AdS_5$ factor, one constructed from two vectors on an $S^2$ factor, and one constructed from one vector from the $AdS_5$ and one from (the same) $S^2$.  We see that there can be no simultaneous zero eigenvalues in the $F$-moduli space.
\item $AdS_3 \times S^2 \times S^2 \times S^2 / H^2 \times S^2 / H^2$ (4.6.5): Taking a generator constructed from two vectors on an $S^2$ factor we see that the generator is skew-symmetric and has constant complex eigenvectors.
\item $AdS_2 \times S^5 / SLAG_3 \times S^2 / H^2 \times S^2 / H^2$ (4.7.12): Taking a generator constructed from two vectors on the $AdS_2$ factor we see that the generator has no zero eigenvalues in the $F$-moduli space.
\item $AdS_2 \times H^5 / (SL(3,\mathbb{R})/SO(3)) \times S^2 \times S^2$ (4.7.12): Taking a generator constructed from two vectors on the $AdS_2$ factor we see that the generator has no zero eigenvalues in the $F$-moduli space.
\item $AdS_2 \times S^3 / H^3 \times S^2 / H^2 \times S^2 \times S^2$ (4.7.14): Taking a generator constructed from two vectors on the $AdS_2$ factor we see that the generator has no zero eigenvalues in the $F$-moduli space via non-flatness of {\em all} $S^2 / H^2$ factors.
\item $AdS_2 \times S^3 \times H^2 \times H^2 \times S^2$ (4.7.14): Taking a generator constructed from two vectors on the $AdS_2$ factor we see that the generator has no zero eigenvalues in the $F$-moduli space via non-flatness of {\em all} $S^2 / H^2$ factors.

\end{itemize}

In all of the above cases we find that all eigenvalues are necessarily nonzero within the $F-$moduli space of the bosonic field equations.  As such, none of the above backgrounds admit supersymmetries.
\end{remark}

\begin{remark}
\label{limitcomp}
Let us consider a geometry $X \times Y$ ($Y$ not flat) and take the geometric limit\cite{geroch1969} in which the curvature of $Y$ goes to zero yielding the limit geometry $X \times T^n$.
For the geometry $X \times Y$ the most general ansatz 4-form $F$ is a sum of available invariant 4-forms whose parameterisation, then constrained by the field equations, forms the solution moduli space.
In this geometric limit, we generally have access to extra invariant forms due to the triviality of the flat component and so must a priori consider extra terms in our 4-form ansatz.
However, we now also have to impose a flatness condition coming from the Einstein equation.

Let us specialise to consider the case where this flatness condition forces all parameters of these extra invariant 4-forms to zero.
In this case the moduli space of the limit geometry is a subspace of the moduli space of our original geometry and the holonomy algebra of the limit geometry is a subalgebra of the holonomy algebra of the original geometry.
As such, we may deduce the absence of supersymmetry of such a limit geometry from the absence of supersymmetry of the original geometry as demonstrated\footnote{As long as we do not use the relevant non-flatness conditions.} via a realised generator common to both holonomy algebras, i.e. constructed from vectors on the $X$ factor.
In this way, we may also rule out supersymmetry for:
\begin{itemize}
\item $AdS_5 \times S^2 \times S^2 \times T^2$ (4.4.3): As a limit of $AdS_5 \times S^2 \times S^2 \times S^2$.
\item $AdS_3 \times S^2 \times S^2 \times S^2 / H^2 \times T^2$ (4.6.5): As a limit of $AdS_3 \times S^2 \times S^2 \times S^2 / H^2 \times S^2$.

\item $AdS_2 \times S^5 / SLAG_3 \times S^2 / H^2 \times T^2$ (4.7.12): As a limit of $AdS_2 \times S^5 / SLAG_3 \times S^2 / H^2 \times S^2$.

\item $AdS_2 \times S^3 / H^3 \times S^2 \times S^2 \times T^2$ (4.7.14): As a limit of $AdS_2 \times S^3 / H^3 \times S^2 \times S^2 \times S^2$.
\item $AdS_2 \times S^3 \times H^2 \times S^2 \times T^2$ (4.7.14): As a limit of $AdS_2 \times S^3 \times H^2 \times S^2 \times S^2$.

\item $AdS_2 \times S^2 / H^2 \times S^2 / H^2 \times S^2 \times T^3$:  As a limit of $AdS_2 \times S^2 / H^2 \times S^2 / H^2 \times S^2 \times S^2 \times S^1$.  The orthogonal symmetry of the $T^3$ component means this limit is the full $F$-moduli space.

\end{itemize}

\end{remark}

\section{Supersymmetric backgrounds}
After these exclusions, we list backgrounds from \cite{jose} which preserve some supersymmetry\footnote{But we do not give supersymmetry moduli-spaces.}:
\begin{itemize}
\item The maximally supersymmetric Minkowski vacuum.
\item Backgrounds of the form $CW_{d>2} \times \R^{11-d}$, where the first factor is a Cahen-Wallach space. These backgrounds are known to be supersymmetric with at least 16 supersymmetries.
\item The well-known maximally supersymmetric Freund-Rubin backgrounds $AdS_7 \times S^4$ and $AdS_4 \times S^7$ (cf. Example \ref{deb}).
\end{itemize}
Other backgrounds from \cite{jose} which preserve some supersymmetry arise as oxidations of well-known lower-dimensional supersymmetric backgrounds. They are:
\begin{itemize}
\item $AdS_3 \times S^3 \times T^5$ and $AdS_3 \times S^3 \times S^3 \times T^2$. Their dimensional reductions $AdS_3 \times S^3 \times T^4$ and $AdS_3 \times S^3 \times S^3 \times S^1$ along $S^1$ to 10d type IIA supergravity are known to admit 16 supersymmetries. Thus, the 11-dimensional geometries admit at least 16 supersymmetries. On the other hand, running the algorithm for these geometries shows directly that there are at most 16 linearly independent spinors annihilated by $\kappa \subset \mathfrak{hol}$.
\item $AdS_{2,3} \times S^{3,2} \times T^6$ admit supersymmetries for $F=f \cdot \sigma_2 \wedge \omega_{T^6}$, where $\sigma_2$ is the area form of the 2-dimensional factor and $ \omega_{T^6}$ denotes the K{\"a}hler form on $T^6$. The dimensional reduction $AdS_{2,3} \times S^{3,2}$ to 5-d supergravity is known to admit 8 supersymmetries. With the same argument as above, also the 11-d geometries admit 8 supersymmetries.
\item $AdS_3 \times S^2 \times S^2 \times T^4$ admits 8 supersymmetries for $F=f_0(\sigma_1 \wedge \sigma_2 + \sqrt{f_1}\sigma_1 \wedge \chi_\text{i} + \sqrt{1-f_1} \sigma_2 \wedge \chi_\text{r})$ where $\sigma_{1,2}$ are the 2-sphere area-forms and $\chi_\text{r,i}$ the real and imaginary parts of the holomorphic 2-form of the $T^4$.
\item $AdS_2 \times S^3 \times S^2 / H^2 \times T^4$ admits 8 supersymmetries for $F=f_0(\nu \wedge \sigma + \sqrt{f_1} \sigma \wedge \chi_\text{r} + \sqrt{1+f_1} \nu \wedge \chi_\text{i})$ where $\nu$ is the $AdS$ area-form, $\sigma$ the 2-sphere area form, and $\chi_\text{r,i}$ the real and imaginary parts of the holomorphic 2-form of the $T^4$
\item $AdS_2 \times S^2 \times S^2 \times T^5$ admits 8 supersymmetries for $F=f_0(\nu \wedge (f_2 d\theta^{34} + f_3 d\theta^{35} - d\theta^{12}) - \sqrt{f_1} \sigma_1 \wedge (f_2 d\theta^{24} + f_3 d\theta^{25} + d\theta^{13}) - \sqrt{1-f_1} \sigma_2 \wedge (f_2 d\theta^{14} + f_3 d\theta^{15} - d\theta^{23}))$ where $\nu$ is the $AdS$ area-form.
\item $AdS_2 \times S^2 \times T^7$ admits 8 supersymmetries for $F=f_0(\nu \wedge \chi_\text{r} \pm \sigma \wedge \chi_\text{i}) + f_1(\nu \wedge \chi_\text{i} \mp \sigma \wedge \chi_\text{r})$ where $\nu$ is the $AdS$ area-form, $\sigma$ the 2-sphere area form, and $\chi_\text{r,i}$ the real and imaginary parts of the holomorphic 2-form of a $T^4$ inside the $T^7$.
\end{itemize}

\section*{Acknowledgments}
We would like to thank Jos\'{e} Figueroa-O'Farrill for initiating this programme and for many fruitful conversations contributing to this result.  We would also like to thank Linus Wulff for pointing out a class of backgrounds missing from an earlier version of this paper.

NH was supported in part by the grant ST/J000329/1 ``Particle Theory at the Tait Institute'' from the UK Science and Technology Facilities Council.

AL acknowledges support from the German Academic Exchange Service (DAAD) and the Collaborative Research Center 647 ``Space-Time-Matter'' of the German Research Foundation.

\begin{appendices}
\section{Conventions}
We list our spinor- and Clifford algebra conventions where we mostly follow \cite{ba81}.\\
Given $p,q \geq 0$ and $n:=p+q$ we denote by $\R^{p,q}$ the vector space $\R^n$ equipped with the inner product
\begin{align*}
\langle x,y \rangle = - \sum_{i=1}^p x_iy_i + \sum_{j=p+1}^n x_jy_j,
\end{align*}
where the coefficients of the vectors $x,y$ are taken wrt. the standard basis $(e_1,...,e_n)$ of $\R^{n}$. The Clifford algebra $Cl(p,q)$ of $\R^{p,q}$ is the up to equivalence unique associative real algebra with unit generated by the relations
\begin{align*}
x \cdot x = -|x|^2.
\end{align*}
Complexification yields a Clifford algebra for $\C^n$. The Clifford algebras $Cl(p,q)$ are isomorphic to standard matrix algebras (or direct sums thereof), cf. \cite{lm}. These isomorphisms allow one to fix irreducible representations
\begin{align}
\rho: Cl(p,q) \rightarrow GL(\Delta_{p,q}), \label{res}
\end{align}
on the real vector space of spinors $\Delta_{p,q}$ which are unique up to equivalence, except that in certain signatures one has to require that the volume element $e_1 \cdot...\cdot e_n$ is mapped to $+1$. Depending on the signature, there are real, quaternionic or complex structures on the space of spinors. 
There is a natural vector space isomorphism
\begin{align*}
Cl(p,q) \cong \Lambda^* \R^{p,q},
\end{align*}
and under this isomorphism the Clifford action of forms on spinors satisfies
\begin{align*}
X^{\flat} \cdot \omega &= X^{\flat} \wedge \omega - X \invneg \omega, \\
\omega \cdot X^{\flat} &= (-1)^p \left(X^{\flat} \wedge \omega + X \invneg \omega \right),
\end{align*}
where $\omega$ is a $p-$form and $X$ a vector.
We denote by $Cl^*(p,q) \subset Cl(p,q)$ the Clifford group of invertible elements. A distinguished Lie subgroup of $Cl(p,q)$ is the spin group
\begin{align*}
Spin(p,q):=\{ x_1 \cdot...\cdot x_{2l} \in Cl_{p,q} : x_j \in \R^{p,q} , \langle x_j , x_j \rangle_{p,q} = \pm 1 \},
\end{align*}
which acts on spinors by restriction of (\ref{res}). The Lie algebra $\mathfrak{spin}(p,q)$ is spanned by the elements $e_i \cdot e_j \in Cl(p,q)$ and we have that $\rho_*:{ \mathfrak{spin}(p,q)} \rightarrow \mathfrak{gl} \left( \Delta_{p,q} \right)$ is given by $\rho_{| \mathfrak{spin}(p,q)}$. Finally, there is a double covering Lie group homomorphism
\begin{align*}
\lambda: Spin(p,q) \rightarrow SO(p,q),\text{ }u \mapsto \left(x \mapsto u \cdot x \cdot u^{-1} \right),
\end{align*}
whose differential is given by $E_{ij} \mapsto 2 e_i \cdot e_j$.\\
\newline
We specialise this discussion to the case of $(p,q)=(10,1)$, giving the real Clifford algebra $Cl(10,1) = M(32,\R) \oplus M(32,\R)$. Thus, the real spinor representation $\rho$ on $\Delta_{10,1}=\R^{32}$ from (\ref{res}) is unique once we specify a value for the volume form. Finally, there is a symplectic structure $\langle \cdot, \cdot \rangle_{\Delta_{10,1}}$ on $\Delta_{10,1}$ which satisfies
\begin{align*}
\langle x \cdot v, w \rangle_{\Delta_{10,1}} = - \langle v, x \cdot w \rangle_{\Delta_{10,1}}.
\end{align*}
In particular, $\langle \cdot, \cdot \rangle_{\Delta_{10,1}}$ is $Spin^+(10,1)$-invariant.
\end{appendices}

\small
\bibliographystyle{utphys}
\bibliography{kssym}

\end{document}